# On Sudden Cessation in Circular Motion

Milan Batista[1]

University of Ljubljana, Slovenia

August 2023

## Abstract

This short paper presents a simple analytical model for the abrupt termination of circular motion, as discussed in the "*The Most Mind-Blowing Aspect of Circular Motion*".[2]

## Introduction

In the web channel program *The Most Mind-Blowing Aspect of Circular Motion* on *All Things Physics*, the presenter examines the circular motion of a ball attached to a string. The central question posed at the beginning of the program is: "*What path does the object take immediately after releasing the string*?" Three possible answers are offered: a) the object continues in circular motion, b) it moves tangentially per circle, and c) it moves normally per circle. Through various experiments, the presenter demonstrates that the correct answer is a) – the object continues to move in a circular motion.

As some commentators have pointed out, it's worth noting that after release, we observe the motion of both the object and the string. If the object detached from the string's end, it would move tangent to the orbit upon release. However, since release occurs from a point on the rotation axis, the motion involves both the object and the string.

For an analytical exploration of this problem, we develop a model using the Lagrange procedure, as detailed in [1]. The model aims to show that the answer a) can be deduced not only through experimentation but also through theoretical means. We aim to provide a theoretical foundation for the observed behaviour by employing the analytical framework presented in this paper.

---

[1] milan.batista@fpp.uni-lj.si

[2] https://youtu.be/AL2Chc6p_Kk

## The model

For the analytical treatment of the problem, we assume that the object is connected to a spring with stiffness $k$, dissipative coefficient $k_d$, and unloaded length $\ell_0$. The object's mass is denoted as $m$, while the mass of the spring is represented as $m_0$. We describe the spring using a lumped model in which half of its mass is attached to its ends. The motion occurs in the O$xy$ plane, with the spring fixed at the coordinate origin. The plane is frictionless, ensuring the object-spring system's motion remains unaffected.

Upon release, the coordinates of the object's and the spring's end are denoted as $(x_1, y_1)$, and the spring's starting point coordinates are at $(x_0, y_0)$. The system's kinetic energy is given by

$$T = \frac{m_0/2}{2}\left(\dot{x}_0^2 + \dot{y}_0^2\right) + \frac{m + m_0/2}{2}\left(\dot{x}_1^2 + \dot{y}_1^2\right), \quad (1)$$

and the potential energy by

$$V = \frac{k(\ell - \ell_0)^2}{2}, \quad (2)$$

where $\ell = \sqrt{(x_1 - x_0)^2 + (y_1 - y_0)^2}$ represents the spring's deformed length. Thus the Lagrange function is

$$L = T - V = \frac{m_0/2}{2}\left(\dot{x}_0^2 + \dot{y}_0^2\right) + \frac{m + m_0/2}{2}\left(\dot{x}_1^2 + \dot{y}_1^2\right) - \frac{k\left(\sqrt{(x_1 - x_0)^2 + (y_1 - y_0)^2} - \ell_0\right)^2}{2}, \quad (3)$$

The Rayleigh dissipation function is taken to be

$$F = \frac{k_d}{2}\left[(\dot{x}_1 - \dot{x}_0)^2 + (\dot{y}_1 - \dot{y}_0)^2\right]. \quad (4)$$

We assume that the spring becomes detached from the centre when it is in the vertical position. Therefore, the initial conditions are as follows:

$$x_0(0) = y_0(0) = x_2(0) = 0, \quad \dot{x}_0(0) = \dot{y}_0(0) = \dot{y}_1(0) = 0, \quad (5)$$

$$y_2(0) = r, \quad \dot{x}_1(0) = \omega_0 r. \quad (6)$$

In Eqs (6), $r$ represents both the initial extension length of the spring and the initial radius of the circle. This length can be obtained by utilizing the dynamic equilibrium equation

$$(m + m_0/2)\omega_0^2 r = k(r - \ell_0) \quad (7)$$

which gives the value

$$r = \frac{k\ell_0}{k-(m+m_0/2)\omega_0^2} = \frac{\ell_0}{1-(m+m_0/2)\omega_0^2/k}. \tag{8}$$

To obtain the dimensionless form of equations, we utilize $\ell_*$ as the unit of length and $\tau_*$ as the unit of time, defined as

$$\ell_* = \ell_0, \quad \tau_* = \sqrt{\frac{m+m_0/2}{k}}. \tag{9}$$

This allows us to express time and coordinates using dimensionless quantities, represented as

$$t = \tau_*\tau, \quad x = \ell_0\xi, \quad y = \ell_0\eta. \tag{10}$$

Here $\tau$ denotes dimensionless time, $\xi$, $\eta$ stands for dimensionless coordinates depend on $\tau$.

The resulting equation of motion, derived from the Lagrangian function as given by Eq (3) and utilizing dimensionless quantities as described in Eqs (10), can now be expressed in the form

$$\left.\begin{aligned}
\ddot{\xi}_0 - \beta(\dot{\xi}_1 - \dot{\xi}_0) - \mu\frac{(\xi_1-\xi_0)\left(\sqrt{(\xi_1-\xi_0)^2+(\eta_1-\eta_0)^2}-1\right)}{\sqrt{(\xi_1-\xi_0)^2+(\eta_1-\eta_0)^2}} &= 0, \\
\ddot{\eta}_0 - \beta(\dot{\eta}_1 - \dot{\eta}_0) - \mu\frac{(\eta_1-\eta_0)\left(\sqrt{(\xi_1-\xi_0)^2+(\eta_1-\eta_0)^2}-1\right)}{\sqrt{(\xi_1-\xi_0)^2+(\eta_1-\eta_0)^2}} &= 0, \\
\ddot{\xi}_1 + \frac{\beta}{\mu}(\dot{\xi}_1 - \dot{\xi}_0) + \frac{(\xi_1-\xi_0)\left(\sqrt{(\xi_1-\xi_0)^2+(\eta_1-\eta_0)^2}-1\right)}{\sqrt{(\xi_1-\xi_0)^2+(\eta_1-\eta_0)^2}} &= 0, \\
\ddot{\eta}_1 + \frac{\beta}{\mu}(\dot{\eta}_1 - \dot{\eta}_0) + \frac{(\eta_1-\eta_0)\left(\sqrt{(\xi_1-\xi_0)^2+(\eta_1-\eta_0)^2}-1\right)}{\sqrt{(\xi_1-\xi_0)^2+(\eta_1-\eta_0)^2}} &= 0.
\end{aligned}\right\} \tag{11}$$

The nonzero initial conditions (6) are

$$\eta_1(0) = \frac{1}{1-\Omega_0^2}, \quad \dot{\xi}_1(0) = \frac{\Omega_0}{1-\Omega_0^2}. \tag{12}$$

In the above equations

$$\mu \equiv 1 + 2\frac{m}{m_0} > 1, \quad \beta \equiv \frac{4k_d\tau_*}{m_0} \geq 0 \quad \text{and} \quad \Omega_0 \equiv \omega_0\tau_*$$

are nondimensional parameters. We assume $m_0 < m$ thus implying $\mu > 1$. Additionally, we consider $k_d \geq 0$ so $\beta \geq 0$, with $\beta = 0$ when there is no dissipation present. Moreover, we assume that $0 \leq \Omega_0 < 1$, signifying that the value of $\omega_0$ must be sufficiently large to extend the string. The case $\Omega_0 = 0$ represents a rigid spring with $k$ approaching infinity, which can be understood as a rigid body.

Before solving the above initial value problem, we look at the motion of the system's centre of mass.. To achieve this, we multiply Eqs (11)₃ and (11)₄ by $\mu$, adding them to Eqs (11)₁ and (11)₂, respectively and then divide the obtained results by $1+\mu$. By integrating these resulting equations while adhering to the initial conditions (12), we derive the motion of the centre of mass, given by

$$\xi_C = \frac{\xi_0 + \mu \xi_1}{1+\mu} = \frac{\mu}{1+\mu}\frac{\Omega_0}{1-\Omega_0^2} \cdot \tau, \quad \eta_C = \frac{\eta_0 + \mu \eta_1}{1+\mu} = \frac{\mu}{1+\mu}\frac{1}{1-\Omega_0^2}. \tag{13}$$

Following the detachment of the string, the centre of mass follows a uniform horizontal straight-line motion, as expected.

## Solution

In general, the equations (11) has no analytical solution and must be therefore integrated numerically. However, given our focus solely on the initial motion, we derive the solution in the form of a power series. The coordinates of interest are those of the object and are given by

$$\xi_1 = \frac{\Omega_0}{1-\Omega_0^2}\tau - \frac{\beta \Omega_0}{\mu(1-\Omega_0^2)}\frac{\tau^2}{2} + O(\tau^3), \quad \eta_1 = \frac{1}{1-\Omega_0^2} - \frac{\Omega_0^2}{1-\Omega_0^2}\frac{\tau^2}{2} + O(\tau^3). \tag{14}$$

By eliminating time from the last two equations, we arrive at

$$\eta_1 = \frac{1}{1-\Omega_0^2} - (1-\Omega_0^2)\frac{\xi^2}{2} + O(\xi^3) \tag{15}$$

As a result, the initial motion follows a parabolic trajectory. For a rigid spring ($k = \infty$), the parabolic trajectory simplifies to

$$\eta_1 = 1 - \tfrac{1}{2}\xi^2 + O(\xi^4). \tag{16}$$

It's worth noting that the parameters $\mu$, and $\beta$ primarily contribute to third and higher-order terms, thus their impact on the initial motion is not substantial. During this initial phase, the key parameter is $\Omega_0$. However, in the special case of a massless string wher $m_0 = 0$, implying $\mu = \infty$, the first two equations within (11) yield $\ell = \ell_0$. This result indicates that in the case of a massless spring, it promptly reverts to its original length. Following this, the third and fourth equations within the set (11) demonstrate that, upon releasing the string, the object moves uniformly in a straight line tangent to the initial circular trajectory.

Finally, let's now address the question of whether the parabola of the initial motion follows the initial circle. The equation for the initial circle is given by:

$$\eta = \sqrt{\frac{1}{(1-\Omega_0^2)^2} - \xi^2} = \frac{1}{1-\Omega_0^2} - \tfrac{1}{2}(1-\Omega_0^2)\xi^2 + O(\xi^2). \tag{17}$$

This match with Eq (15) up to the second order term confirms that the object will indeed follow the circle's path in its initial motion, as claimed in the program.

## Conclusions

In this paper, we've demonstrated that comprising a spring and a point object, the proposed model accurately predicts the start of circular motion immediately after the string is detached. Notably, this successful prediction relies on the spring possessing mass. In cases where the string is massless, the object's motion becomes tangential to the circle. Interestingly, this circular motion doesn't seem to be significantly influenced by how the spring loses energy, indicating that energy loss has a minor effect at this stage.

After the initial circular motion, it's important to note that the object doesn't just follow the tangent of the circle. Instead, it oscillates to a line parallel to the direction the centre of mass is moving. This can be demonstrated by numerically solving the stated initial value problem